\documentclass{jpsj-suppl}
\usepackage{txfonts} 
\usepackage{url}
\usepackage[dvipdfmx]{}
\usepackage{amsmath}
\usepackage{color}

\newcommand{\etal}{{\it et al}.}

\title{Possibility of Systematic Study of Supernova Explosions by Nuclear Imaging Spectroscopy}

\author{
Yoshitaka~\textsc{Mizumura}$^{1,2}$,
Toru~\textsc{Tanimori}$^{2,1}$ and
Atsushi~\textsc{Takada}$^{2}$
on behalf of the SMILE group$^{3}$
}

\inst{
$^{1}$Unit of Synergetic Studies for Space, Kyoto University, Sakyo, Kyoto 606-8502, Japan \\
$^{2}$Graduate School of Science, Kyoto University, Sakyo, Kyoto 606-8502, Japan
$^{3}$\url{http://www-cr.scphys.kyoto-u.ac.jp/research/MeV-gamma/}
}

\email{mizumura@cr.scphys.kyoto-u.ac.jp}

\recdate{August 20, 2016}

\abst{
An all-sky monitor with nuclear imaging spectroscopy is 
a promising tool for the systematic study of supernova explosions.
In particular, progenitor scenarios 
of type-Ia supernovae, which are not yet well understood,
can be resolved using light curves in the nuclear gamma-ray band.
Here we report an expected result of an all-sky monitor with imaging spectroscopy
using electron-tracking Compton camera,
which will enable us to observe nuclear gamma-ray lines from type-Ia supernovae.
}

\kword{type-Ia supernovae, gamma rays, imaging spectroscopy}

\begin{document}
\maketitle

\section{Introduction}

Type-Ia supernovae (SNe) are known to be one of the most interesting objects in astronomy;
nonetheless, their progenitor systems are not yet well understood.
The MeV gamma-ray light curve is expected to be a promising tool
to distinguish between the progenitor scenarios
(i.e., single degenerate (SD) versus double degenerate (DD))\cite{Summa-13}.
Recently, SPI/{\it INTEGRAL} successfully detected $^{56}$Ni and $^{56}$Co
decay nuclear gamma-ray lines from SN2014J\cite{Diehl-14,Churazov-14},
which is the closest type-Ia SN (at a distance of 3.5~Mpc) observed in the past four decades.
Although SPI has a large effective area (several tens of cm$^{2}$)\cite{Attie-03},
it is difficult to resolve the progenitor scenario from the obtained light curves\cite{Churazov-15}.
In addition, HXD/{\it Suzaku} also attempted to measure soft gamma rays from SN2014J;
the signal-like excesses with about $2\sigma$ confidence level
and $3\sigma$ upper limit flux has been reported\cite{Terada-16}.
Both SPI and HXD suffer from intense background radiations\cite{Diehl-14,Terada-16}.
%
We have developed an electron-tracking Compton camera (ETCC),
which is the first MeV gamma-ray telescope to provide a well-defined point spread function (PSF)
within a few degrees and efficient background rejection\cite{Takada-11,Tanimori-15}.
We note the PSF is defined with two directional angles similar to optics,
whereas in standard Compton cameras, the angular resolution is defined with only one directional angle.
We have revealed a satellite equipped with four modules of (50~cm)$^3$-sized ETCCs,
that will reach a sensitivity of 1~mCrab\cite{Tanimori-15}.
Here we present the expected MeV observations of SNe
using nuclear imaging spectroscopy realized using the sharp PSF.

\section{Extragalactic MeV gamma-ray background}

\begin{figure}[ht]
  \begin{center}
  \includegraphics[width=0.7\textwidth]{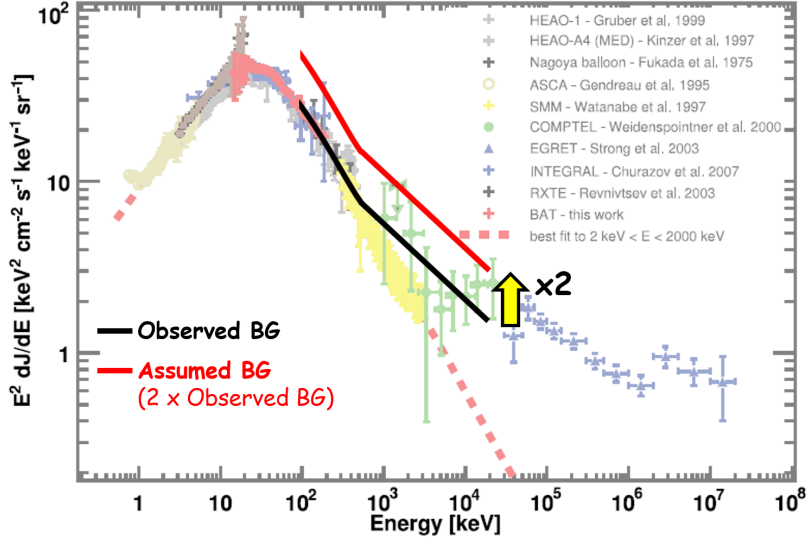}
  \caption{
    Extragalactic diffuse background spectra compiled by the BAT/{\it Swift} team\cite{Ajello-08}.
    We assumed the background-radiation level, as shown by the solid red line,
    which is two times higher than the observed background level indicated by the black line.
    Details concerning the black line are reported in the text.
  }
  \label{f1}
  \end{center}
\end{figure}

From the observations of {\it INTEGRAL} and {\it Suzaku},
we find that estimations of the background level have a serious effect on MeV gamma-ray astronomy.
Background origins for gamma-ray telescopes can be categorized into five groups;
diffuse gamma-ray radiations,
non-gamma-ray particles (i.e., charged particles and neutrons),
incomplete gamma-ray events (i.e., escaped events of Compton-recoil electrons and/or Compton-scattered gamma rays),
chance coincidence hits,
and the major background are gamma rays emitted from the detector and satellite itself.
A well-defined PSF is the most effective tool for background suppression similar to optical/X-ray telescopes,
which suppresses for the most part of gamma rays coming from the outside of the PSF in particular major components,
and it enables us to estimate the remaining contamination quantitatively.
A particle identification using energy-loss rate (dE/dx) selects events in which the electron stopped in the gaseous volume\cite{Tanimori-15}.
The dE/dx rejects non-gamma-ray particles and electron-escaped events.
Also kinematics test using $\alpha$ angle suppresses events which not-satisfied to Compton kinematics,
such as chance coincidence hits, gamma-ray-escaped events, and Compton-scattering events at the scintillators.
Combination of all of the above background-rejection methods are possible only for ETCC.
The inefficiencies of background rejection with dE/dx and $\alpha$-selection
for even an early stage ETCC in 2006 without a well-defined PSF was reported to be
$(1.8\pm0.3)\times10^{-5}$ for neutrons,
$(3.4\pm0.2)\times10^{-4}$ for electrons, and
$(2.5\pm0.2)\times10^{-4}$ for protons\cite{Takada-11}.
Figure~\ref{f1} shows the extragalactic diffuse gamma-ray spectra compiled by
the BAT/{\it Swift} team\cite{Ajello-08}.
The spectra can be fit well using a smoothed broken power law in the energy ranges lower than 500~keV.
However, the COMPTEL data\cite{Weidenspointner-00} has excesses
in the MeV band in comparison to the fitted results.
To take these excesses into consideration,
we assumed the following conditional equation
(the black line in Fig.~\ref{f1})
as observed in the extragalactic diffuse gamma-ray spectrum (hereafter, the {\it observed-BG}):
\begin{eqnarray}
E^{2} \frac{{\rm d}J}{{\rm d}E} =
\left\{
  \begin{array}{ll}
E^{2} \times \frac{\displaystyle C_1}{\displaystyle \left( E/E_{brk} \right)^{\Gamma_1} + \left( E/E_{brk} \right)^{\Gamma_2} } & (E \leq 500~{\rm keV}), \\
E^{2} \times C_2 \left( \frac{\displaystyle E}{\rm 500~keV} \right)^{-\Gamma_3} & (E > 500~{\rm keV}),
  \end{array}
\right.
\label{eq1}
\end{eqnarray}
where $E$ is the photon energy,
$E_{brk} = 29.99$~keV is the broken energy,
$C_1 = 10.15\times10^{-2}$ and $C_2 = 30.32\times10^{-6}$ are the normalization factors,
and $\Gamma_1 = 1.32$, $\Gamma_2 = 2.88$, and $\Gamma_3 = 2.44$ are the power-law indices,
respectively.
In addition, we conservatively assumed a two times higher intensity than in Eq.~(\ref{eq1})
for the background radiation level (the solid red line in Fig.~\ref{f1};
hereafter,
the {\it assumed-BG}).
The conservative-factor of two includes contribution of background of non-gamma-ray origins,
which of contributions due to instrumental gamma-ray emissions and inefficiencies of background rejections
are 19\% and 2\% level of diffuse gamma-ray radiations, respectively\cite{Takada-11}.
Therefore, totally about 21\% level of diffuse gamma-ray radiations is sufficiently less than the conservative-factor.
The {\it assumed-BG}
is used in the estimation of MeV observations of type-Ia SNe with the satellite ETCCs.

%

\section{Observable gamma-ray spectra from type-Ia SNe}

\begin{figure}[ht]
  \begin{center}
  \includegraphics[width=0.9\textwidth]{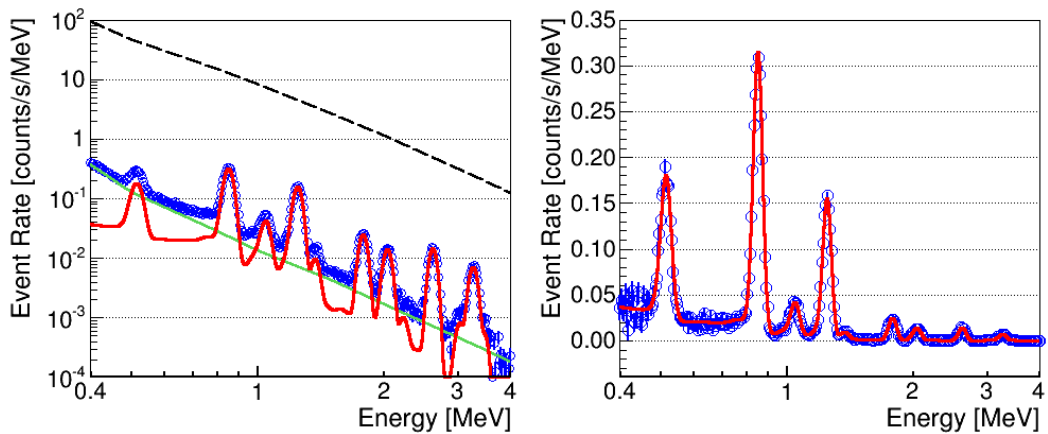}
  \caption{
    Expected gamma-ray spectrum using observations by the satellite ETCCs
    in 50--100~days following an explosion.
    The solid red line indicates the estimated gamma-ray rate from an SD-scenario type-Ia SN
    that exploded at a distance of 5~Mpc.
    The solid green and dashed black lines indicate the contamination of the {\it assumed-BG}
    within the sharp PSF of the satellite ETCCs, and half of the {\it observed-BG}, respectively.
    The open blue circles in the left and right panels show
    the expected total gamma-ray spectrum within the sharp PSF
    and the residual gamma-ray spectrum minus an equivalent level of the {\it assumed-BG} contamination,
    respectively.
  }
  \label{f2}
  \end{center}
\end{figure}

We estimated gamma-ray observations of type-Ia SNe with the satellite ETCCs.
Details concerning the configuration and sensitivity of the modules
are described in Tanimori~\etal\cite{Tanimori-15}.
As source spectra, 
we used the gamma-ray spectra of type-Ia SNe calculated by Summa~\etal\cite{Summa-13}.
The time evolution of the source spectra during 100~days following the explosion was considered.
The background spectrum was calculated by
the combination of the background-radiation level, effective area, and PSF of the satellite ETCCs.
Figure~\ref{f2} shows a typical observable gamma-ray spectrum from a type-Ia SN and the background levels.
The solid-red line indicates the estimated gamma-ray rate from an SD-scenario type-Ia SN,
that exploded at a distance of 5~Mpc.
The solid green and dashed black lines indicate the contamination of the {\it assumed-BG}
within the sharp PSF of the satellite ETCCs, and half of the {\it observed-BG}, respectively.
Here we assumed an energy resolution of $\Delta E/E = 5.0\times(E/662~{\rm keV})^{-0.5}~[\%]$ (FWHM)
as the detector response.
The energy resolutions of the ETCCs are assumed to be typical of high-resolution scintillators
(e.g., GAGG\cite{Iwanowska-13} and LaBr$_3$\cite{Quarati-11})
that are photo-absorber candidates for the satellite ETCCs.
The open blue circles in Fig.~\ref{f2} show
the expected total gamma-ray spectrum within the sharp PSF (left panel),
and the residual gamma-ray spectrum minus an equivalent level of the {\it assumed-BG} contamination (right panel),
which is estimated using the off-source sky
in the wide field of view of the ETCCs ($2\pi$~str\cite{Tanimori-15}).
The {\it observed-BG} level is 10--100 times more intense than the source spectrum.
From this study, we expect the detection of several nuclear lines of $^{56}$Co
using observations of the satellite ETCCs.

\section{Conclusions}
We have estimated observations of nuclear gamma-ray lines
from a SD-scenario type-Ia SN exploded at a distance of 5~Mpc using the ETCC technology.
We assumed a moderate energy resolution for the detector;
however, the result clearly suggests that
the satellite ETCCs will enable us to study nuclear gamma-ray lines of
SD-scenario type-Ia SN at distances beyond SN2014J ($3.5$~Mpc).
Therefore, we believe that the imaging suppression of the background-radiation level based on the PSF
is a significant component for gamma-ray spectroscopy of type-Ia SNe.
The cases of more distant SNe and DD-scenario SNe are also important for systematic studies of SNe.
This is beyond the scope of this short paper; however, we are preparing to publish a study of these cases.

\end{document}